\documentclass{eptcs}
\providecommand{\event}{G.Ciobanu, M.Koutny (Eds.): Membrane Computing and\\
Biologically Inspired Process Calculi (MeCBIC 2010), 2010.}

\usepackage{amsmath}
\usepackage{amssymb}
\usepackage{amsthm}
\usepackage{amsfonts}
\usepackage{graphicx}

\usepackage{BioPEPA}

\newcommand{\ltrans}[1]{\xrightarrow{#1}}

\newcommand{\agr}{\quad\big|\quad}

\newcommand{\cN}{{\cal N}}
\newcommand{\cK}{{\cal K}}
\newcommand{\cP}{{\cal P}}
\newcommand{\cD}{{\cal D}}
\newcommand{\cPt}{\tilde{\cP}}
\newcommand{\cPtt}{\overline{\cP}}

\newcommand{\defP}{\stackrel{\mathit{def}}{=}}

\newenvironment{definition}[1][Definition]{\begin{trivlist}
\item[\hskip \labelsep {\bfseries #1}]}{\end{trivlist}}

\newcommand{\Sfrom}[1]{ ((\alpha, \kappa){#1}S)(l, L)}
\newcommand{\Sto}[2]{ (S)(#1, #2)}
\newcommand{\System}[1]{ \langle {\cal V}, {\cal N}, {\cal K}, {\cal F}, Comp, {#1} \rangle}
\newcommand{\Systemd}[1]{ \langle {\cal V}, {\cal N}, {\cal K}, {\cal F}, Comp, \sigma,  {#1} \rangle}

\newcommand{\capab}[1]{ \;{\ltrans{(#1)}_{st}}\; }
\newcommand{\defcapab}[1]{ \alpha^+, [S:{#1}(l,\kappa)] }
\newcommand{\defmul}[1]{  L @ [(l ,\kappa, \alpha, {#1})]}

\newcommand{\compl}[1]{ \;{\ltrans{(#1)}_{co}}\; }
\newcommand{\defcompl}[1]{ \alpha^-, [S:{#1}(l,\kappa)] }

\newcommand{\stoch}[1]{ \;{\ltrans{(#1)}_{s}}\; }

\title{Modeling biological systems with delays in  Bio-PEPA}

\author{Giulio Caravagna
\institute{Dipartimento di Informatica,\\ Universit\`a di Pisa,\\Largo Pontecorvo 3, 56127 Pisa, Italy.}
\email{caravagn@di.unipi.it}
\and
Jane Hillston
\institute{Laboratory for Foundations of Computer Science,\\ The University of Edinburgh,\\ Edinburgh EH8 9AB, Scotland.}
\email{jane.hillston@ed.ac.uk}
}
\def\titlerunning{Modeling biological systems with delays in  Bio-PEPA}
\def\authorrunning{G. Caravagna and J. Hillston}

\date{}
\begin{document}
  \maketitle 
  
\begin{abstract}
Delays in biological systems may be used to model events for which the underlying dynamics cannot be precisely observed, or to provide abstraction of some behavior of the system resulting more compact models.  In this paper we enrich the stochastic process algebra Bio-PEPA, with the possibility of assigning delays to actions, yielding a new non-Markovian process algebra: Bio-PEPAd.   This is a conservative extension meaning that the original syntax of Bio-PEPA is retained and the delay specification which can now be associated with actions may be added to existing Bio-PEPA models. The semantics of the firing of the actions with delays is the delay-as-duration approach, earlier presented in papers on the stochastic simulation of biological systems with delays.  These semantics of the algebra are given in the Starting-Terminating style,  meaning that the state and the completion of an action are observed as two separate events, as required by delays.  Furthermore we outline how to perform stochastic simulation of Bio-PEPAd systems and how to automatically translate a Bio-PEPAd system into a set of Delay Differential Equations, the deterministic framework for modeling of biological systems with delays.  We end the paper with two example models of biological systems with delays to illustrate the approach.
\end{abstract}

\section{Introduction}

The contribution of computer science in the interdisciplinary field of Systems Biology is to provide languages, tools and techniques for the description and analysis of complex biological systems. In particular, there exist many formal languages, either based on process algebras or term--rewriting systems, worth noting: Bio-PEPA \cite{biopepa1, biopepa2}, the stochastic $\pi$-calculus \cite{priami, spi2, spi3}, Bioambients \cite{bioambients}, the $\kappa$-calculus \cite{kappa}, LBS \cite{lbs}, the CLS \cite{cls1,cls2,cls3}, to name but a few.

Biological systems can often be modeled at different abstraction levels. Specifically, a
simple event in a model that describes the system at a certain level of detail
may correspond to a rather complex network of events in a lower level
description. The choice of the abstraction level of a model usually depends on
the knowledge of the system and on the efficiency of the analysis tools to be
applied to the model. 

Delays can appear in a biological system at any level of abstraction. In particular, there are two good reasons for considering delays. Firstly, when there is a network of events whose dynamics cannot be precisely observed, or measured in terms of kinetic rates and, secondly, when a complex portion of a system is to be abstracted by means of a smaller one. In both cases, a delay may represent the time necessary for the underlying network of
events to produce some result observable in the higher level model. In both cases, the state space of the model with delays will be a reduction of the complete one as some parts are abstracted by the delays.

In mathematics, the modeling of  biological systems with delays is mainly based on
Delay Differential Equations (DDEs), a kind of differential equations, obtained by generalizing Ordinary Differential Equations (ODEs), in which
the derivative of the unknown function at a certain time is given in terms of
the values of the function at previous times. This framework is very general and allows both simple (constant) and complex (variable or distributed) forms of delays to be modeled. Practically, DDEs have been used to describe biological systems in which events have a non-negligible duration \cite{BERETTA,ZHANGA} or in which a sequence of simple events is abstracted as a single complex event associated with a duration \cite{VR03,HIV}.

It is well-known that the analysis of ODEs can become imprecise due to the approximation introduced by representing discrete quantities with continuous variables when quantities are close to zero, and the same problem can arise in DDEs. Thus techniques for performing stochastic simulation of biological systems with delays have been defined. The Delay Stochastic Simulation Algorithms (DSSAs) \cite{io,io2,barrio}, often exploiting the  Gillespie's Stochastic Simulation Algorithm (SSA) of chemical reactions \cite{gillespie}, permit computation of a time--trace of the non--Markovian stochastic process underlying a model with delays. These algorithms permit different interpretation of delays \cite{io, io2}, in particular, it is possible to have a \emph{delay-as-duration} approach to the firing of reactions, or a \emph{purely delayed} one. In the former \cite{io, barrio}, the reactants are removed at the beginning of a reaction and the products are added at its end, namely after the delay plus an exponentially distributed time quantity. In this sense, during the time of firing of the reaction, the reactants will not be able to take part in other reactions. However, in \cite{io,io2} the need for a different interpretation of delays is discussed via an example of the cell--cycle with delays. More precisely, it is shown that, for some biological systems, it is necessary that reactants involved in a reaction with delay can have other interactions while waiting for the delay to complete. Indeed, the latter interpretation, namely the purely delayed approach, is such that the reactants involved in a reaction can have other interactions during the firing of the reaction itself. The reactions are hence scheduled and fully performed after the delay and the stochastic time quantity have expired, if the reaction is still applicable.

In this paper, we define a process algebra for the modeling of biological systems with delays. More precisely, we use constant delays in the DDEs and, for the DSSAs, we take the delay-as-duration approach presented in \cite{io}. These restrictions are reasonable since they permit us to have a simple algebra obtained by extending a well-known one, Bio-PEPA \cite{biopepa1,biopepa2}. Also, later versions of this algebra may be extended to more complex forms of delay and interpretation of delays.

Bio-PEPA is a stochastic process algebra for the modeling and the analysis of biochemical networks. Bio-PEPA  is based on PEPA \cite{pepa}, a process algebra originally defined for the performance analysis of computer systems, and extends it in order to handle some features of biochemical networks, such as stoichiometry and different kinds of kinetic laws. A main feature of Bio-PEPA is the ability to support different kinds of analysis. In particular, Bio-PEPA models can be analyzed by performing stochastic simulation based on the Gillespie's SSA \cite{gillespie} and steady state analysis can be performed on the Continuous--Time Markov Chain underlying the semantics of a model. Furthermore, Bio-PEPA models can be translated into equivalent deterministic models based on ODEs and, finally, they can be model checked using the PRISM \cite{prism2, prism} model checker. The Bio-PEPA modeling paradigm is {\em processes-as-species} rather than {\em processes-as-molecules}, as in the Stochastic $\pi-$calculus \cite{priami}. This choice, in general, permits a smaller state space and hence a model whose analysis is feasible. 

In this paper, we enrich the stochastic process algebra Bio-PEPA with the possibility of assigning delays to actions, yielding the definition of a new non--Markovian process algebra: Bio-PEPAd. The new algebra is based on the same syntax as Bio-PEPA, hence the definition of Bio-PEPAd systems with delays can be easily obtained by adding, to a Bio-PEPA system of the target model, the delay specifications. Bio-PEPAd contains two issues to tackle model reduction: the use of the level of concentrations for the species, as in Bio-PEPA, and the delays, as a new feature. The semantics of the algebra is given in the Starting-Terminating style \cite{bravetti}, which allows us to observe the start and the completion of an action as two separate events, as required by delays. Following previous work on Bio-PEPA analysis,  we outline how to perform stochastic simulation of Bio-PEPAd systems using the DSSAs introduced in \cite{io}, and how to automatically translate a Bio-PEPAd system in a set of  DDEs. 

At the end of the paper we present some examples of biological systems described by Bio-PEPAd. In particular, we show the semantics of a toy model in order to clarify the ideas underlying the definition of the algebra. Also, we encode in Bio-PEPAd a well-known model of the cell--cycle with delays where the passage of cells from different phases of the cell cycle is modeled by a delay. Such a model is then translated into a set of DDEs which match with the first definition of the model, appearing in \cite{VR03}. We end the paper with some discussions about the future work we plan. 

The paper is structured as follows: in Section \ref{sect:biopepa} we recall the definitions of Bio-PEPA that we maintain in the definition of Bio-PEPAd. In Section \ref{sect:biopepad} we separately introduce the syntax and the semantics of the language. In Section \ref{sect:analysis} we present analysis techniques for Bio-PEPAd systems based on DDEs and DSSAs. In Section \ref{sect:examples} some examples of Bio-PEPAd systems are presented and, finally, in Section \ref{sect:conclusions} conclusions and future work are discussed.

\section{Bio-PEPA} \label{sect:biopepa}

Bio-PEPA \cite{biopepa1,biopepa2} is a stochastic process algebra, based on  PEPA \cite{pepa},  for the modeling and the analysis of biochemical networks. The operators of this algebra are designed for easily describing biochemical networks. Indeed, features such as stoichiometry of reactions and general kinetic laws can be easily described in Bio-PEPA models. Furthermore, as already said in the previous section, the algebra supports multiple analysis techniques for the defined models. Stochastic simulations, steady state analysis of the CTMC, automatic translation in sets of deterministic ODEs and, finally, model checking analysis can be performed on Bio-PEPA models.

The processes-as-species modeling paradigm of Bio-PEPA permits a smaller state space and, consequently, a model whose analysis is feasible. A model is described by sequential components representing species, and by some model components representing their possible interactions. 

In this section we recall the parts of the definition of Bio-PEPA that we will use to define Bio-PEPA with delays. We assume a
set of action types $\cA$ and we start by recalling the syntax of the processes.
\begin{definition}
Bio-PEPA processes are defined by the following grammar:
\begin{align*} 
S \; &::= \; (\alpha, \kappa) op\; S \agr S + S \agr C \\
P \; &::= \; P \sync{\cal L} P  \agr S(l)
 \end{align*}
where $op = \reactant \mid \product \mid \modifier \mid \activator \mid \inhibitor$, $\alpha \in \cA$, $\cal L$ is a set of actions and $l, \kappa \in \mathbb{N}$. 
We denote with $\cS$ the set of all possible species specifications, and we denote with $\cP$ the set of all possible well--formed Bio-PEPA processes, as defined in \cite{biopepa1}.
\end{definition}
The components $S$ and $P$ represent species and their possible interactions, respectively. The element $C$ is used to define constant processes. 

Bio-PEPA actions are used to model the events (i.e. the reactions) happening in the biological systems we model. The prefix terms in this algebra contain information about the role of the species in the actions. In particular, for $ (\alpha, \kappa) op\; S$ we have that $(\alpha, \kappa)$ is the prefix, where $\alpha \in \cA$ is the action type and $\kappa$ is the stoichiometry 
coefficient of the species in the reaction. The prefix combinator {\em ``op''} represents the role of the species in the reaction. In particular, $\reactant$ indicates a reactant, $\product$ a product, $\activator$ an activator, $\inhibitor$ an inhibitor and 
$\modifier$ a generic modifier.  The actions can appear in a summation term $S_1+S_2$, whose meaning is the classical ``choice" of process algebras.

In Bio-PEPA a discrete concentration level $l$ is associated with each species. During the simulation of a system, the concentration of  a species $S$, denoted by $S(l)$ ranges over $\{0, \ldots, N_S\}$, where $N_S$ is its maximum level of concentration statically defined to bound the population size. Also, a fixed step size $h$ for all the species is defined. This means that, changing the level concentration of a species by one, implies a change in $h$ units of concentration of that species. The granularity, as well as the rate functions, are defined in terms of the step size $h$ of the  concentration intervals.  This choice permits us to deal with incomplete information in the exact number of elements, and leads to a reduction of the state space as there are less states for each component. 

Bio-PEPA supports multiway synchronization, i.e. synchronization can involve more than two components. This makes it easy to model n-ary reactions, whose modeling in dyadic process algebras is not trivial. The term $P_1 \sync{\cal L} P_2$ denotes cooperation between $P_1$ and $P_2$ over the \emph{cooperation set} ${\cal L}$, which determines those activities on which the cooperands are  forced to synchronise. For action 
types not in $\cal L$, the components proceed independently and concurrently with 
their enabled activities. 

A Bio-PEPA model specification is given in terms of systems, where a system is defined as follows.
\begin{definition}
A Bio-PEPA system $\cP$ is a 6-tuple $\System{P}$ where: 
\begin{itemize}
\item ${\cal V}$ is the set of compartments;
\item ${\cal{ N}}$ is the set of quantities describing each species;
\item ${\cal K}$ is the set of parameter definitions;
\item ${\cal {F}}$ is the set of functional rate definitions;
\item ${Comp}$ is the set of sequential component definitions;
{\item $P$ is  the initial process definition.}
\end{itemize}
\end{definition}

{ Notice that in Bio-PEPA the kinetic characteristics of the actions are not specified in the syntax of processes as in other calculi but, instead, they are separately represented in the notation of system. Indeed, in this definition the information about  rates  is given by $\cal{F}$ and that about kinetic constants is given by $\cal{K}$, while the initial process definition is $P$.}

The semantics of Bio-PEPA is given by a Structural Operational Semantics (SOS) \cite{plotkin}, similar to the one for PEPA. 
The semantics is based on a capability relation which supports the derivation of quantitative information and which is auxiliary to a stochastic relation. The stochastic relation associates the rates with the actions performed. 
The rates are obtained by evaluating the functional rate associated with the action, divided by the step size, and by using the 
quantitative information derived from the capability relation, as explained in \cite{biopepa1}. The use of two relations allows for the association of the rate with the last step of the 
derivation representing a given reaction, which makes it easier to derive the rate in 
the appropriate way, especially in the case of general kinetic laws different from 
mass-action. 

For the precise definitions and explanations of the components of a Bio-PEPA system, as well as for the formal definition of the SOS of Bio-PEPA, we refer to \cite{biopepa1}.

\section{Bio-PEPAd: Bio-PEPA with delays} \label{sect:biopepad}

In the following sections we separately present the syntax and the semantics of Bio-PEPA with delays (Bio-PEPAd).

\subsection{Syntax and process configurations}
Processes of Bio-PEPAd are defined by the same syntax as Bio-PEPA processes, hence
it will be possible to easily encode a Bio-PEPA system in one with delays. 

{As in Bio-PEPA the general kinetic information is specified separately from the syntax of processes. The delays, which are also properties of the actions which can be performed, are similarly represented separately  in Bio-PEPAd. Indeed, they are defined by functions as }
\begin{align}\label{def:del}
\sigma: \cA \mapsto \{ r \in \mathbb{R} \mid r \geq 0\} 
\end{align}
such that $\sigma(\alpha)$ denotes the delay of action $\alpha \in \cA$. {From the biological perspective, the choice of using $\sigma$ to specify the delays implies that, for every participant in an action $\alpha$, a unique delay $\sigma(\alpha)$ corresponds, which is sound since for each species involved in the reaction modeled by $\alpha$ the delay is unique.} A Bio-PEPAd system is defined as an extension of a Bio-PEPA one as follows.
\begin{definition}
A Bio-PEPAd system is  a 7-uple $\Systemd{P}$ where: 
\begin{itemize}
\item $\System{P}$ is a Bio-PEPA system;
\item $\sigma$ is a function satisfying (\ref{def:del}) and used to specify the delays of the actions.
\end{itemize}
We denote with $\cPt$ the set of all possible Bio-PEPAd systems. 
\end{definition}
Again, moving from a Bio-PEPA system specification to a Bio-PEPAd one is straightforward. This will permit us, in the future, to reuse the system specifications for Bio-PEPA in the context of Bio-PEPAd. In order to define the semantics of Bio-PEPAd we define a notion of process configuration.
\begin{definition}
Bio-PEPAd process configurations are defined by the following syntax:
\begin{align*} 
C_S \; &::= \; (\alpha, \kappa) op\; C_S \agr C_S + C_S \agr C \\
C_P \; &::= \; C_P \sync{\cal L} C_P  \agr C_S(l, L)
 \end{align*}
where $L$ is a list of 4-tuples $(l, \kappa, \alpha, op)$ with $l, \kappa \in \mathbb{N}$, $\alpha \in \cA$ and $op = \reactant \mid \product \mid \modifier \mid \activator \mid \inhibitor$. We denote with $\cC$ the set of all well--formed processes configurations. 
\end{definition}

The notion of well--formed process configuration is straightforward; any process configuration is well--formed if, by removing the list $L$, its corresponding Bio-PEPA process is well--formed. For clarity, in the following we denote a generic process configuration as $S(l,L)$.

A species $S(l, L)$ is a species with a discrete level of concentration $l$, like the species $S(l)$ in Bio-PEPA, but which is currently involved in the actions with delay described by the list $L$. In particular, if the list $L$ contains an entry $(l', \kappa, \alpha, op)$, this means that there are $\kappa$ levels of concentration of species $S$ involved in a currently running action $\alpha$ which fired when the discrete level of concentration of species $S$ was $l'$, its role in this instance of action is described by $op$.

Consequently, $L$ is to be considered as a view of the scheduling list used in the algorithms  described in \cite{io} for simulating stochastic models with delays. More precisely, $L$ is a view of only the scheduled events which involve elements of species $S$. 

In order to define the semantics of Bio-PEPAd, it is necessary to define some auxiliary functions for manipulating the scheduling list $L$. We denote with $\cD$ the domain of all the possible tuples of the form $(l, \kappa, \alpha, op)$, and with ${\cal L_D}$ all the possible lists built over $\cD$, hence $L \in {\cal L_D}$.
We start by defining, in a functional style \cite{ml}, a function $\phi: \cA \mapsto {\cal L_D} \mapsto \cD$ to extract the first scheduled events with a given action name from the list $L$ as follows:
\begin{align*}
\phi \; \alpha \;L \;&= \; \underline{match} \;\; L\;\; \underline{with} \\
& | \; [\;] \; \to \; \perp; \\
& | \; (l, \kappa, \alpha, op)::xs \; \to \; (l, \kappa, \alpha, op); \\
& | \; x::xs \; \to \;  \phi \; \alpha \;  xs.
\end{align*}
The function value $\phi \; \alpha \; L$ is $\perp$ if no entries of action $\alpha$ exist in $L$ (i.e. no actions $\alpha$ are currently running), otherwise it is the first entry obtained by a left-to-right  recursive scan of $L$. Notice that we assume the syntactic priority of pattern matching.

Now, we  define a function $\zeta: \cA \mapsto {\cal L_D} \mapsto {\cal L_D}$ such that $\zeta \; \alpha \; L$ is a new list obtained by removing the first, if any, occurrence of an action $\alpha$ obtained by a  left-to-right  recursive scan of $L$. We define $\zeta$  as follows:
\begin{align*}
\zeta \; \alpha \;L \;&= \; \underline{match} \;\; L\;\; \underline{with} \\
& | \; [\;] \; \to \; [\;]; \\
& | \; (l, \kappa, \alpha, op)::xs \; \to \; xs; \\
& | \; x::xs \; \to \;  x :: \zeta \; \alpha \;  xs.
\end{align*}
As this is an event list, the ordering of insertion of the tuples determines their ordering for extraction. The functions $\phi$ and $\zeta$, together with the classical append function on lists, namely function $@$, will be used  to implement a First-In First-Out (FIFO)  policy for insertion and extraction of elements in $L$.

Furthermore, as in Bio-PEPA we want to keep the state representation of the models finite by using some constraints for the starting of actions. Thus, let us denote the scheduled actions in which the species $S$ is involved as a product by $\pi \; L$, where $\pi:  {\cal L_D} \mapsto {\cal L_D}$ is a recursive function defined as
\begin{align*}
\pi \; L \;&= \; \underline{match} \;\; L\;\; \underline{with} \\
& | \; [\;] \; \to \; [ \; ]; \\
& | \; (l, \kappa, \alpha, \product)::xs \; \to \; (l, \kappa, \alpha, \product) :: \pi \; xs; \\
& | \; x::xs \; \to \;  \pi \;   xs.
\end{align*}
The species $S(l,L)$ is currently involved in the delayed actions as follows: for the scheduled actions in $\pi \; L$ it is involved  as a product, and for the other ones it is involved either as a reactant, a modifier, an activator or an inhibitor.
Furthermore, let us denote by $\rho:  {\cal L_D} \mapsto \mathbb{N}$ the function
\begin{align*}
\rho \; L \;&= \; \underline{match} \;\; L\;\; \underline{with} \\
& | \; [\;] \; \to \;0;\\
& | \; (l, \kappa, \alpha, op)::xs \; \to \; \kappa + \rho \; xs.
\end{align*}
This function computes how many levels of concentration are involved in all the actions described in its input list, regardless of the role of the species in the scheduled event. By following the delay-as-duration of approach \cite{io} in the interpretation of the delays this implies that, for species $S$, there are exactly $\rho \; \pi \; L$ levels of concentrations of species $S$ which are currently waiting for their delay to expire before becoming available in the species $S$. These two functions will be used to define the constraints to keep the state space finite, as presented in the next sections.

A Bio-PEPAd system specification is typically given in terms of a process $P \in \cP$ 
whose semantics is given in terms of its equivalent process configuration $P_C \in \cC$. Intuitively, we want the initial term $P$ to be modified in the corresponding initial configuration $P_C$ where every species declaration $S(l_{0}, [ \; ])$ in $P_C$ is such that $S(l_0)$ is in $P$. The initial process
configuration is obtained by adding an empty scheduling list  to each species because, in the initial configuration, there are no instances of actions with delay currently running. Formally, we define, by structural recursion on the processes' syntax a function $\mu: \cP \mapsto \cC$ such that
\begin{align*}
&\mu ((\alpha,\kappa) op\; S) = (\alpha, \kappa) op\; S &\mu( P_1 \sync{\cal L} P_2) = \mu(P_1)   \sync{\cal L} \mu(P_2)\\
& \mu(S_1+S_2) = S_1+S_2  &\mu(S(l)) = S(l, [ \; ]).
\end{align*}

We augment the definition of Bio-PEPAd systems to 7-tuples of the form $\Systemd{P_C}$ where $P_C$ is a process configuration of a process. In the following, we may use the notation $P$ to refer to either a process or a process configuration; it will be clear from the context to which of them we are referring.
We denote the extended set of all Bio-PEPAd systems with process configurations as $\cPtt$.

Similarly to Bio-PEPA where the SOS is defined by means of two relations, in this algebra the SOS, given in a Starting--Terminating (ST) style \cite{bravetti}, is defined by means of three relations that we present in the following section. 

\subsection{A Structural Operational Semantics}

In the following subsections we define a {\em start relation} on process configurations which, in the same style as the Bio-PEPA capability one, contains the quantitative information needed to evaluate the functional rates and modifies the process configurations to model the start of an action. Also, we define a {\em completion relation} on process configuration which describes the termination of an action. Finally, along the line of the stochastic relation in Bio-PEPA, we define a {\em stochastic relation} for Bio-PEPAd systems, based on the start and completion relations,  which associates rates with transitions.

\subsubsection*{The start relation}

This relation contains the quantitative information to compute rates of starting actions. Also, this relation modifies the process configuration to model the starting of an action. 

The start relation is $\ltrans{}_{st} \subseteq \cC \times \Theta^+ \times \cC$ where $\theta ^+ \in \Theta^+$ contains, like the capability relation in Bio-PEPA, the information to evaluate the functional rate. We define the labels $\theta^+$ as
\[
\theta^+ := (\alpha^+, w) 
\]
where $w$ is defined as $w ::= [S: op(l, \kappa)] \mid w @ w$, with $S \in \cS$, $op$ an operator, $l$ the level and $\kappa$ the stoichiometry coefficient of the components. The list $w$ is  of the same type as the one exhibited as a label by the capability relation of Bio-PEPA. The Bio-PEPAd start relation is defined as the minimum relation satisfying the rules presented in Figure \ref{fig:capability}.

\begin{figure}[t]
\centering
\begin{gather*} 
\Sfrom{\reactant} \capab{\defcapab{\reactant}} \Sto{l-\kappa}{\defmul{\reactant}} \qquad \kappa \leq l \leq N\\
 \Sfrom{\product\;} \capab{\defcapab{\product}} \Sto{l}{\defmul{\product}} \qquad   0 \leq l + \rho \; \pi\; L \leq N \\
 \Sfrom{op\;} \capab{\defcapab{op}} \Sto{l}{\defmul{op}} \qquad  1 \leq l \leq N \; {\text and}\; op=\modifier, \inhibitor \\
 \Sfrom{op\;} \capab{\defcapab{op}} \Sto{l}{\defmul{op}} \qquad \kappa \leq l \leq N \; {\text and}\; op=\activator  \\
 \frac{S_1(l, L) \capab{\alpha^+, w} S_1'(l', L')}{(S_1+ S_2) (l, L) \capab{\alpha^+, w} S_1'(l', L')}\qquad \quad
 \frac{S_2(l, L) \capab{\alpha^+, w} S_2'(l', L')}{(S_1+ S_2) (l, L) \capab{\alpha^+, w} S_2'(l', L')}\\
 \frac{P_1 \capab{\alpha^+, w} P_1' \quad \alpha \not \in {\cal L}}{P_1\sync{\cal L} P_2 \capab{\alpha^+, w} P_1'} \qquad \quad
 \frac{P_2 \capab{\alpha^+, w} P_2' \quad \alpha \not \in {\cal L}}{P_1\sync{\cal L} P_2 \capab{\alpha^+, w} P_2'} \qquad \quad
 \frac{P_1 \capab{\alpha^+, w_1} P_1' \quad P_2 \capab{\alpha^+, w_2} P_2' \quad \alpha \in {\cal L}}{P_1\sync{\cal L} P_2 \capab{\alpha^+, w_1 @ w_2} P_1' \sync{\cal L} P_2' } 
 \end{gather*}
\caption{The start relation $\ltrans{}_{st} \subseteq \cC \times \Theta^+ \times \cC$.}
\label{fig:capability}
\end{figure}

Formally, if a species $\Sfrom{\reactant}$ is involved as reactant in an action, then by following the delay-as-duration approach \cite{io} its concentration level is decreased by $\kappa$.  Differently, in the case of a species involved as a product, its concentration level is not changed because, as previously stated, this relation models the starting and not the completion of an action with delay. In the case of a species taking part in the reaction as a modifier,  an inhibitor or an activator, its concentration level is not changed, as expected. Independently of the role of a species, its scheduling list $L$ is modified to record that some of its levels of concentration are currently performing action $\alpha$. Notice that, in order to maintain the FIFO property on the scheduling list $L$, we simply use the append function $@$. This is possible because of both the multiway synchronization of Bio-PEPAd and the use of fixed deterministic delays. More precisely, two instances of the same action starting in two subsequent instants, are assumed to terminate in two  subsequent instants. This is true in a framework where delays are deterministic however, if they were stochastic, the two instances could have multiple orderings for completion.  Indeed, because of the multiway synchronization in the scheduling list $L$ the two instances will appear subsequently and, hence, will complete subsequently. Notice that, in a process calculus with dyadic synchronization, this would not have held by simply using function $@$.

We use constraints on the levels to have a finite state space as in Bio-PEPA.
The constraints for starting the actions are the same as those in Bio-PEPA except the one for the products. In particular, the constraints which must be satisfied by a species $S(l,L)$ to fire an  action  as a product is, as expected, $0 \leq l + \rho \; \pi \; L \leq N$, if $N$ is its maximum level. Intuitively, this means that the levels of concentration in the state, $l$, plus those which already scheduled to be produced, $\rho \; \pi \; L$, must not cross the capacity threshold $N$.

The starting of the action $\alpha$, in the style of the ST semantics \cite{bravetti}, is denoted by the action symbol $\alpha^+$, exhibited as a label for all the start derivations. The composition of the derivations of this relation is straightforward.

Some further considerations and comparisons with Bio-PEPA are useful. Firstly, when the actions have no delay as in Bio-PEPA,  whenever an action fires, the changes in the process are immediately visible in a one--step derivation, since the Bio-PEPA capability relation modifies the process according to the action. In this algebra, as the instant in which an action starts and terminates are detached, then the start relation modifies the process  to represent just the starting of the action. Indeed, another relation, which does not exist in the semantics of Bio-PEPA, will model the termination of a currently running action. 

Secondly, by comparing the algorithm presented in \cite{io} and the definition of this relation, it is clear that the modification of the process to reflect the starting of an action corresponds to scheduling of the reaction in the scheduling list. 
\subsubsection*{The completion relation}

This relation is used to model the completion of an action with delay which is currently running. Also, this relation contains quantitative information needed to re--compute the functional rate of the action at the moment in which it started.

The completion relation is $\ltrans{}_{co} \subseteq \cC \times \Theta^- \times \cC$ where $\theta ^- \in \Theta^-$ is a similar label to the one exhibited by the start relation. We define the labels $\theta^-$ as
\[
\theta^- := (\alpha^-, w) 
\]
where $w$ is defined as for the start relation. The completion relation is defined as the minimum relation satisfying the rules of Figure \ref{fig:completion}.

\begin{figure}[t]
\centering
\begin{gather*}
\frac{\phi \; \alpha \; L = (l, \kappa, \alpha, \product)}{
S(l',L) \compl{\defcompl{\product}} S(l'+k, \zeta \; \alpha \; L)  }\qquad
\frac{\phi \; \alpha \; L = (l, \kappa, \alpha, op)}
{S(l', L) \compl{\defcompl{op}} S(l', \zeta \; \alpha \; L)} \quad \;op=\reactant, \modifier, \activator, \inhibitor\\
 \frac{S_1(l, L) \compl{\alpha^-, w} S_1'(l', L')}{(S_1+ S_2) (l, L) \compl{\alpha^-, w} S_1'(l', L')}\qquad \quad
 \frac{S_2(l, L) \compl{\alpha^-, w} S_2'(l', L')}{(S_1+ S_2) (l, L) \compl{\alpha^-, w} S_2'(l', L')}\\
 \frac{P_1 \compl{\alpha^-, w} P_1' \quad \alpha \not \in {\cal L}}{P_1\sync{\cal L} P_2 \compl{\alpha^-, w} P_1'} \qquad \quad
 \frac{P_2 \compl{\alpha^-, w} P_2' \quad \alpha \not \in {\cal L}}{P_1\sync{\cal L} P_2 \compl{\alpha^-, w} P_2'} \qquad \quad
 \frac{P_1 \compl{\alpha^-, w_1} P_1' \quad P_2 \compl{\alpha^-, w_2} P_2' \quad \alpha \in {\cal L}}{P_1\sync{\cal L} P_2 \compl{\alpha^-, w_1 @ w_2} P_1' 1\sync{\cal L} P_2'}
  \end{gather*}
\caption{The completion relation $\ltrans{}_{co} \subseteq \cC \times \Theta^- \times \cC$.}
\label{fig:completion}
\end{figure}

Formally, for a species $S(l,L)$ it is possible to get the instance of a currently running action $\alpha$, if any,  by applying function $\phi$. More precisely, this permits us to get, from all the possible instances of actions $\alpha$, the first which has been scheduled, $\phi \; \alpha \; L$, and, hence, the first which will terminate. If the species is involved as a product, then it is necessary to increase, as defined by the delay-as-duration approach,  its concentration level by adding the scheduled products. Otherwise, whatever the role of the species, its concentration level must remain constant. 
Independently of the role of the species in the action, the scheduling list is modified by means of the function $\zeta$, hence a new list $\zeta \; \alpha\; L$ is produced by removing from $L$ the entry which was computed by function $\phi$.

Obviously, no constraints are stated for the completion of a currently running action, as the appropriate ones are checked before the starting of actions.

The completion of the action $\alpha$, in the style of the ST semantics \cite{bravetti}, is denoted by the action  symbol $\alpha^-$, exhibited as a label for all the completion derivations. The other label, namely the list $w$, is defined like the one exhibited by the start relation. The composition of this relation with the other operators is straightforward and very similar to the composition of the derivations of the start relation. 

Some further considerations are worth noting. Firstly, this relation is a new one with respect to the Bio-PEPA semantics. Again, in a framework where actions have no delays the contribution of this relation to the semantics would have been given by means of a unique relation. Also, as the role of this relation is to model the completion of an action, it chooses actions to terminate from the list which is associated with the species, namely the list of actions currently running. The start relation, differently, chooses the action to fire from the species definition. 

Furthermore, as we want the completion relation to exhibit quantitative information to recompute the functional rate of the action at the moment at which it started, then the labels exhibited by this relation are very similar to those exhibited by the start relation, even
if they are computed starting from $\phi \; \alpha \; L$. This permits us to have a unique policy for computing the functional rates from the input lists, obtained by derivations of the transitions of these relations.

\subsubsection*{The stochastic relation}

The stochastic relation permits us to associate rates to transitions. Also, this transition permits us to observe changes in a Bio-PEPAd system  due to either the starting or the completion of an action. 

The stochastic relation is $\stoch{} \subseteq \cPtt \times \Gamma \times \cPtt$ where $\gamma \in \Gamma$ is defined as $$\gamma := (\alpha^+, r_\alpha,{\sigma}_\alpha) \mid (\alpha^-, r_\alpha, {\sigma}_\alpha)$$ with $\alpha \in \cA$, $r_\alpha \in \mathbb{R}^+$ and ${\sigma}_\alpha \in \mathbb{R}$. As in Bio-PEPA, $r_\alpha$ represents the parameter of an exponential distribution and, as expected, all activities enabled attempt to proceed but only the fastest succeeds. 

As this relation is defined on the set $\cPtt$, namely the set 
of all possible Bio-PEPAd systems with process configurations, whenever we
refer to the semantics of a system $\Systemd{P}$, where $P$ is a process, we assume we apply the stochastic relation to the system $\Systemd{\mu(P)}$. Again, this is necessary because $P$ is not a process configuration, and we want to build, from $P$, the corresponding initial configuration $\mu(P)$, and then we want apply the semantics to the system.

The stochastic relation is defined as the minimum relation satisfying the  rules given in Figure \ref{fig:stochastic}.

\begin{figure}[t]
\centering
\begin{align*}
&\frac{P \capab{\alpha^+, w} P'}{\Systemd{P} \stoch{\alpha^+, r_\alpha[w, {\cal N}, {\cal K}], \sigma(\alpha)} \Systemd{P'}} \\
&\frac{P \compl{\alpha^-, w} P'}{\Systemd{P} \stoch{\alpha^-, r_\alpha[w, {\cal N}, {\cal K}], \sigma(\alpha)} \Systemd{P'}} 
 \end{align*}
\caption{The stochastic relation $\ltrans{}_{s} \subseteq \cPtt \times \Gamma \times \cPtt$.}
\label{fig:stochastic}
\end{figure}

Formally, the starting of an action $\alpha$, obtained by composition with a derivation of the start relation, is denoted by symbol $\alpha^+$. The completion of an action is obtained by composition with a derivation of the completion relation, as denoted by symbol $\alpha^-$. 

The rate of any action is computed as in Bio-PEPA, namely as $r_\alpha [w, \cN, \cK] ={f_\alpha[w, \cN, \cK]} \cdot {h^{-1}}$. For the explanation of how the rates are computed because of the levels we refer to \cite{biopepa1}. For any possible derivation of the stochastic relation, the value $\sigma(\alpha)$ denotes the delay of the action $\alpha$.

A {\em Stochastic Labelled Transition Systems} can be defined for a Bio-PEPA system with delays.
\begin{definition} The {\em Stochastic Labelled Transition Systems} (SLTS) for a Bio-PEPAd system is $(\cPtt, \Gamma, \ltrans{}_{s})$ where $\ltrans{}_{s}$ is the minimal relation satisfying the rules given in Figure \ref{fig:stochastic}.
\end{definition}

\section{Analysis techniques}\label{sect:analysis}

In this section we present some analysis techniques for Bio-PEPAd systems along the line of those presented in \cite{biopepa1}  for Bio-PEPA systems. Firstly, we discuss the automatic translation of a Bio-PEPAd system into a set of Delay Differential Equations (DDEs). Secondly, we discuss how to apply a Delay Stochastic Simulation Algorithm (DSSA) to compute the stochastic time--evolution of a Bio-PEPAd model.

\subsection{Translation in Delay Differential Equations} \label{sect:ddes}

Whenever phenomena presenting a delayed effect are described by differential equations, we move from ODEs to DDEs. In DDEs the derivatives at current time depend on some past states of the system. The simplest form of DDE considers {constant delays}, namely consists of
equations of the form
\[
\frac{dX}{dt}=f_x(t,X(t),X(t-\sigma_1),\ldots,X(t-\sigma_n))
\]
with $ \sigma_1>\ldots>\sigma_n\geq 0$, $\sigma_i \in \mathbb{R}$ and $X(t-\sigma_i)$ denotes the state of the system at the past time $t-\sigma_i$. This form
of DDE allows models to describe events which have a fixed duration. Hence it is natural, in the context of Bio-PEPAd, to reason about the translation of a model into a set of DDEs. Furthermore, similar work has been presented in \cite{biopepa1} for translating a Bio-PEPA system into a  set of ODEs.

In order to define the encoding it is important to recall that we defined Bio-PEPAd in terms of Bio-PEPA. This means that, given a system specification $\Systemd{P}$ where $P$ is a valid Bio-PEPA process, we just need to modify the algorithm defined in \cite{biopepa1} to add the information provided by $\sigma$ concerning the delays. Formally, the results for Bio-PEPA permit us to encode $\System{P}$ in a set of ODEs by using the definition of the stoichiometry matrix associated with $P$.

The algorithm presented in \cite{biopepa1} consists of three steps. In the first the stoichiometry matrix is defined, in the second the kinetic law vector $\nu_{KL}$ is derived and in step three the deterministic variables are associated with the components. Steps $(1)$ and $(3)$ are unaffected by the use of delays; hence we preserve them.

Step $(2)$, namely the definition of the kinetic law vector $\nu_{KL}$, must be changed. Such a vector contains the kinetic law of each reaction; we will explain the definition via an example. For instance, for an action $\alpha$ involving species $S_1$ and $S_2$, and with mass action kinetics $f_{MA}(k)$, its original entry in the vector $\nu_{KL}$ would be $kx_{S_1}(t)x_{S_2}(t)$ where $x_{S_1}$ and $x_{S_2}$ are the deterministic variables representing species $S_1$ and $S_2$ species, respectively. The variables depend on the state at time $t$, but in the context of DDEs, the delays of the actions become dependencies on the past states of the system. Hence, for that particular example, the correct entry in the vector $\nu_{KL}$ must be $kx_{S_1}(t-\sigma(\alpha))x_{S_2}(t-\sigma(\alpha))$. Step $(2)$ can be generalized adding, in the process of the definition of $\nu_{K_L}$, the delays of the form $x_S(t - \sigma(\alpha))$ for all species $S$ and actions $\alpha$.

The DDE system can be defined in the same way as the ODE one, namely as ${d\overline{x}}/{dt} = D \times \nu_{KL}$ where $\overline{x}$ and $D$ are the results of step $(3)$ and $(1)$ of the algorithm, respectively. The initial conditions are, however, different from the ones defined for ODEs. In particular, the DDEs, because of the delays, must be defined also in the interval $[t_0-{\sigma(\alpha)}; t_0]$ where $\alpha$ is the action with maximum delay. 

It is not possible to define a universal initial condition for the DDEs systems as every possible configuration will affect the dynamics of the whole system. Sometimes the initial conditions of a species $S$ are defined via a constant function $\phi_S(t) \mbox{ for } t \in[t_0-{\sigma(\alpha)}; t_0]$ such that $\phi_S(t) = hl_{S,0}$ where $l_{S,0}$ is the initial concentration level for $S$ in the Bio-PEPAd model and $h$ is the step size for the concentration levels {(see \cite{io} for details)}.
In general, we leave this part of the translation to the modeler who will tune the initial conditions with respect to the specification of the target system.

\subsection{Stochastic Simulation}  \label{sect:dssa}
The stochastic simulation of biological systems is typically based on the SSA by Gillespie \cite{gillespie} and its variants. Anyway, the SSA, as well as all its variants, are not able to deal with actions with delays but only with Markovian actions. As a consequence, some DSSAs \cite{io,barrio}  have been defined to perform the stochastic simulation of a system where actions have a fixed delay. 

In this section we briefly explain how to perform the stochastic simulation of a Bio-PEPAd system by using the DSSAs presented in \cite{io}, where all the reactions follow a delay-as-duration approach. Also in this context the choice of reusing part of the Bio-PEPA definitions in BIo-PEPAd is crucial. In particular, this permits us to completely re-use the techniques defined in \cite{biopepa1} to perform the stochastic simulation of Bio-PEPA systems by using the SSA.

The main steps in preparing a Bio-PEPAd system for the application of the DSSA are the two. Firstly, given an initial system $\Systemd{P}$, where $P$ is a Bio-PEPA process as well as  a Bio-PEPAd one, a  vector describing the initial numbers of molecules to be simulated must be obtained by an encoding of $P$. Secondly, the actual rates of the reactions have to be defined, by cases, starting from $\mathcal{F}$.

Both the two issues are independent of the target algorithm (hence of the delays). More precisely, since both the SSA and the DSSA simply assume a vector $\mathbf{X}(t_0)$ describing the initial state of the system, and as $P$ is also a Bio-PEPA process since it is not in a configuration, then  we can simply use the techniques developed in  \cite{biopepa1} to compute $\mathbf{X}(t_0)$ from $P$.

Also the definition of the actual rates of the reactions can be done again in the same way for both Bio-PEPA and Bio-PEPAd systems. This is possible since  the DSSA we refer to is based on the SSA. Indeed, it assumes the same type of definition of propensity functions to compute the probability of the reactions,. Finally, as $\mathcal{F}$ is defined in the same way for both Bio-PEPA and Bio-PEPAd, then the techniques developed in \cite{biopepa1} to define from $\mathcal{F}$ the actual rates can be used also in the context of Bio-PEPAd.

Once these two steps have been performed, the resulting systems can be simulated by the DSSA presented in \cite{io,io2} where all the actions follow a delay-as-duration approach.

\section{Examples}\label{sect:examples}

In this section we present a toy example of a Bio-PEPAd model to illustrate the approach, followed by an encoding of a well--known model of the cell cycle in Bio-PEPAd.

\subsection*{A toy example}

In order to clarify the modeling with  Bio-PEPAd we present a toy example of a model. We assume a single reaction channel of the form $A \xrightarrow{k, \sigma'} B$ to denote a transformation of an element of species $A$
into an element of species $B$ with a kinetic constant $k$ and a delay $\sigma'$. The initial state contains three elements of species $A$ and no elements of species $B$; formally it is described by the vector $\mathbf{X}(t_0) = (3, 0)^T$.

The Bio-PEPAd processes modeling the species can be easily defined as follows:
\[
A \defP (\alpha, 1) \reactant A \qquad \quad B \defP (\alpha, 1) \product B
\]
where $\alpha$ is the action which models the reaction, the functional rates are defined according to the mass--action kinetics, $f_\alpha= f_{MA}(k)$ and the delay
is defined according to the function $\sigma(\alpha) = \sigma'$. The Bio-PEPAd process describing the interacting components is $A \sync{\{\alpha\}}B$. By considering levels we assume the species to have some maximum levels $N_A$ and $N_B$ where $N_A > 3$ and $N_B>3$. The initial levels of concentrations are described by the vector $\mathbf{X}(t_0)$, and the initial configuration of the process, obtained by applying function $\mu$, is the following
\[
A (3, [\;]) \sync{\{\alpha\}}B (0, [\;])\, .
\]

By applying the stochastic relation to the system with this process configuration we obtain all the possible evolutions of the configuration. The obtained LTS, as expected, is finite, and, because of the delays, it corresponds to a non--Markovian stochastic process. Intuitively, there is a one-to-one correspondence between both the states and the transitions of the LTS and of the stochastic process, exactly as between the LTS for Bio-PEPA systems and the CTMCs.

\begin{figure}[t]
\begin{center}
\includegraphics[scale=0.6]{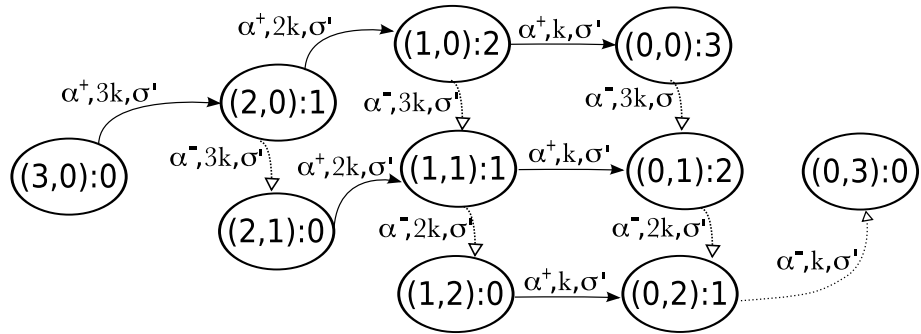}
\caption{A graphical representation of the stochastic process obtained by applying the semantics to the process configuration for the toy example.}
\label{fig:toy}
\end{center}
\end{figure}

A graphical representation of the stochastic process is given in Figure \ref{fig:toy}. In that figure, all the states are represented as circles where the notation $(n_1, n_2):m$ represents the discrete levels of concentration of the species $A$, $n_1$, and $B$, $n_2$. The number $m$ represents the number of instances of the unique possible action $\alpha$ currently scheduled in the state. All the arrows represent stochastic derivations of the whole system, where the labels are exactly those computed by that relation. The full arrows represent stochastic derivations based on start derivation, empty arrows  represent stochastic derivations based on completion derivation. For this particular example, any empty arrow built from a derivation with a rate $r$ refers to the completion of the unique action started with the same rate $r$. 

Figure \ref{tab:toy} presents a table showing the explicit mapping of the states described in Figure \ref{fig:toy} and the corresponding process configuration obtained by the semantics. For the sake of clarity, as in this simple example there is just one action, $\alpha$, and $A$ always participates in that action as a reactant and $B$ as a product, this information is omitted from the scheduling lists.

As expected, this system, starting from the initial configuration $\mathbf{X}(t_0)$, namely state$(3,0):0$, eventually reaches the final state $(0,3):0$, which corresponds to the final configuration $A (0, [\;]) \sync{\{\alpha\}}B (3, [\;])$ and to the vector $\mathbf{X}(t')=(0,3)^T$, for some $t' > t_0$.

\begin{figure}[t]
{\small
\begin{center}
  \begin{tabular}{| c || c |}
    \hline
    state & process configuration \\ \hline
    $(3,0):0$ & $A (3, [\;]) \sync{\{\alpha\}}B (0, [\;])$ \\ \hline
    $(2,0):1$ & $A (3, [(3,1)]) \sync{\{\alpha\}}B (0, [(0,1)])$  \\ \hline
    $(2,1):0$ & $A (2, [\;]) \sync{\{\alpha\}}B (1, [\;])$ \\ \hline
    $(1,0):2$ & $A (1, [(3,1),(2,1)]) \sync{\{\alpha\}}B (0, [(0,1),(0,1)])$ \\ \hline
    $(1,1):1$ & $A (1, [(2,1)]) \sync{\{\alpha\}}B (1, [(1,1)])$ \\ \hline
    $(1,2):0$ & $A (1, [\;]) \sync{\{\alpha\}}B (2, [\;])$ \\ \hline
    $(0,0):3$ & $A (1, [(3,1),(2,1),(1,1)]) \sync{\{\alpha\}}B (0, [(0,1),(0,1),(0,1)])$ \\ \hline
    $(0,1):2$ & $A (1, [(2,1),(1,1)]) \sync{\{\alpha\}}B (1, [(1,1),(1,1)])$  \\ \hline
    $(0,2):1$ & $A (0, [(1,1)]) \sync{\{\alpha\}}B (2, [(2,1)])$ \\ \hline
    $(0,3):0$ & $A (0, [\;]) \sync{\{\alpha\}}B (3, [\;])$ \\ \hline
  \end{tabular}
  \caption{A table stating the correspondence between the states represented in Figure \ref{fig:toy} and the process configurations obtained by the semantics.}
  \label{tab:toy}
\end{center}
}
\end{figure}


\subsection*{A model of the cell cycle with delays}

In this section we encode  in Bio-PEPAd a model of the cell cycle with delays as presented in \cite{io}. Such a model is obtained by simplifying a  DDEs model of tumor growth that includes the immune system response and a phase-specific drug able to alter the natural course of action of the cell cycle of the tumor cells \cite{VR03} . 

The model of the cell cycle with delays has been analyzed in \cite{io} in order to discuss two possible interpretations of delays in the delay stochastic simulation algorithms, a delay-as-duration approach and a purely delayed approach. In this section, we simply show how to encode that model  in Bio-PEPAd and for a detailed analysis of the model we refer to that paper.

The cell cycle is a series of sequential events leading to cell replication via cell division. It consists of four phases: G$_1$, S, G$_2$ and M. The first three phases (G$_1$, S, G$_2$) are called interphase. In these phases, the main event which happens is the replication of DNA.
In the last phase (M), called mitosis, the cell segregates the duplicated sets of chromosomes between daughter cells and then divides to form two new cells in their interphase. The duration of the cell cycle  depends on the type of cell (e.g a normal human  cell takes approximately 24 hours to perform a cycle). Cell death via apoptosis may happen in any phase of the cell cycle.

The Bio-PEPAd model considers two populations of cells: $T_I$, the population of tumor cells during cell cycle interphase, and $T_M$, the population of tumor cells during mitosis. We consider four possible actions, $\alpha$, $\beta$, $\gamma$ and $\delta$, one for each of the events that we want to model. In particular, action $\alpha$ models the passage from the interphase to the mitotic phase, with rate $a_1$, $\beta$ models the mitosis, with rate $a_4$, $\gamma$ the death of a cell in the interphase, with rate $d_2$, and $\delta$ the death of a cell in the mitotic phase, with rate $d_3$. All the rates in the model refer to mass action kinetics. 

The Bio-PEPAd model is defined by the following species definitions:
\begin{align*}
T_I & \defP (\alpha, 1) \reactant + (\beta, 2) \product + (\gamma, 1) \reactant \\
T_M & \defP (\alpha,1) \product + (\beta, 1) \reactant +  (\delta, 1) \reactant
\end{align*}
where the species behave as reactants or products, depending on their role as previously specified.  Also, as all the actions obey a mass action kinetic law, we simply assume $f_\alpha= f_{MA}(a_1)$, $f_\beta= f_{MA}(a_4)$, $f_\gamma= f_{MA}(d_2)$ and $f_\delta= f_{MA}(d_3)$. The Bio-PEPAd process  modeling the interactions is given by
\[
T_I( n^I_0) \sync{\{\alpha,\beta\}} T_M(n^M_0)
\]
where $n^I_0$ and $n^M_0$ represent the initial concentration levels for the cells in the interphase and in the mitotic phase, respectively. Notice that $\gamma$ and $\delta$ are not in the cooperation set since model reactions involving a single species. Also, we recall that this is also a valid Bio-PEPA process specification.

A delay $\sigma'$ is used to model the duration of the interphase, hence the passage of a tumor cell from the population of those in the interphase to the population of those in the mitotic phase, namely the event modeled by action $\alpha$, is delayed. To specify the delay in the Bio-PEPAd system to analyze,  it is enough to define a function $\sigma$ where 
\begin{align*}
&\sigma(\alpha) = \sigma'  & \sigma(\beta) = \sigma(\gamma) = \sigma(\delta) = 0.
\end{align*}
As a consequence, the Bio-PEPAd process initialized by applying function $\mu$, namely the process configuration $T_I( n^I_0,[\;]) \sync{\{\alpha,\beta\}} T_M(n^M_0, [\;])$, together with the function $\sigma$, completes the definition of the Bio-PEPAd system representing the cell cycle model. 

By applying one of the techniques discussed in this paper this system can be analyzed. 
In particular, the Bio-PEPAd model can be automatically translated into a set of DDEs by applying the algorithm presented in Section \ref{sect:ddes}. By computing the following vector of the kinetic laws
\begin{align*}
\nu_{KL} &= (a_1T_I(t-\sigma(\alpha)), \; a_4 T_M(t -\sigma(\beta)), \; d_2T_I(t-\sigma(\gamma)), \; d_3T_M(t-\sigma(\delta)) )^T \\
& = (a_1T_I(t-\sigma'), \; a_4 T_M(t), \; d_2T_I(t), \; d_3T_M(t) )^T
\end{align*}
the following set of DDEs can be computed:
\begin{align*}
  &\frac{dT_I}{dt}  = 2a_4T_M - d_2 T_I - a_1 T_I(t-\sigma') &
  \frac{dT_M}{dt}  = a_1T_I(t-\sigma') - d_3 T_M - a_4 T_M \, .
\end{align*}
As expected, this DDEs system is analogous to the one presented in \cite{io}.  The terms $d_2T_I$ and $d_3 T_M$ represent  cell deaths. The cells reside in the interphase at least $\sigma'$ units of time; then the number of cells that enter mitosis at time $t$ depends on the number of cells that entered the interphase $\sigma'$ units of time before. This is modeled by the terms $T_I (t - \sigma')$ in the DDEs. Also, each cell leaving the mitotic phase produces two new cells in the $T_I$ population, as given by terms $-a_4T_M$ and  $2a_4T_M$. As a consequence, by defining the appropriate initial conditions for the resulting DDEs system it would be possible to reproduce the results presented in \cite{io} for the deterministic model. 

As far as the stochastic analysis of the Bio-PEPAd systems is concerned, we can notice that the system we defined corresponds to the following set of reactions 
\begin{align*}
T_I \xrightarrow{a_1} T_M \mbox{ with delay $\sigma$} \qquad T_M \xrightarrow{a_4}
2T_I \qquad
T_I \xrightarrow{d_2} \qquad T_M \xrightarrow{d_3}\, .
\end{align*}
Again, this is exactly the same reactions--based model used in \cite{io} to compare the deterministic and the stochastic models for the cell cycle. Consequently, by applying the DSSA as explained in Section \ref{sect:dssa}, it would be possible to reproduce the results presented in \cite{io} for the stochastic model.

\section{Discussion and conclusions} \label{sect:conclusions}

In this paper, we have enriched the stochastic process algebra Bio-PEPA with the possibility of assigning delays to actions, yielding the definition of a new non--Markovian process algebra: Bio-PEPAd.  The use of delays in biological systems is suitable to model events for which the underlying dynamics cannot be precisely observed. Also, delays can be used to abstract portions of systems, leading to a reduced state space for models. From this point of view Bio-PEPA, which is based on the idea of levels to tackle the problem of state space explosion, was an appropriate candidate for defining our algebra.

The algebra is based on the syntax of Bio-PEPA. Hence the definition of Bio-PEPAd systems with delays can be easily obtained by adding, to a Bio-PEPA system of the target model, the delay specifications. 

The semantics of the firing for the actions with delays is the delay-as-duration approach, as presented in the definition of DSSAs. In future work, we may enrich Bio-PEPAd with the other interpretation of delays presented in \cite{io, io2}, in order to have the purely delayed approach and its combination with the one we currently consider.

The semantics of the algebra has been given in the Starting-Terminating style. This permits us to observe the start and the completion of an action as two separate events, as required by delays. In future work, we will consider equivalence relations for Bio-PEPAd systems and processes, as done in \cite{vashti} for the Bio-PEPA ones.

In keeping with the techniques developed for analyzing Bio-PEPA models,  we outlined how to perform stochastic simulation of Bio-PEPAd systems and how to automatically translate a Bio-PEPAd system in a set of  Delay Differential Equations, the deterministic framework for the modeling of biological systems with delays. Moreover, the software framework for Bio-PEPA \cite{biopepaw} could be extended to provide a tool for the automatic analysis of Bio-PEPAd systems.

As a proof of concept, we presented two examples of Bio-PEPAd systems. The first one, a toy example, has been shown to illustrate the semantics we defined. The second one, a well--known model of the cell--cycle where phase passages are abstracted by means of a delay, has been presented in order to show the translation of the Bio-PEPAd system into both a stochastic process with delays to be simulated by a DSSA, and a set of DDEs which can automatically derived by the system specification.

{In the future, we plan to define Bio-PEPAd models of biological systems with delays and to analyze such models using the anlysis techniques we defined in this paper.}

An interesting area for further future work will be to compare Bio-PEPAd with non--Markovian Stochastic Petri Nets such as DSPN \cite{OSPN}.

\bibliographystyle{eptcs} 

\end{document}